\documentclass[12pt]{article}
\usepackage{graphicx}

\input{epsf}
\usepackage{cite}
\def\@fmsl@sh#1#2#3{\m@th\ooalign{$\hfil#1\mkern#2/\hfil$\crcr$#1#3$}}
 \def\eq#1\en{\begin{equation}#1\end{equation}}
\def\s[#1,#2]{[#1\stackrel{\star}{,}#2]}
\def\sx[#1,#2]{[#1\stackrel{\star_{x}}{,}#2]}

\newcommand{\nc}{\newcommand}
\nc{\beq}{\begin{equation}}
\nc{\eeq}{\end{equation}}
\nc{\beqa}{\begin{eqnarray}}
\nc{\eeqa}{\end{eqnarray}}

\def\gsim{\mathrel{\rlap{\lower4pt\hbox{\hskip1pt$\sim$}}
    \raise1pt\hbox{$>$}}}       

\begin{document}
\makeatletter
\def\fmslash{\@ifnextchar[{\fmsl@sh}{\fmsl@sh[0mu]}}
\def\fmsl@sh[#1]#2{%
  \mathchoice
    {\@fmsl@sh\displaystyle{#1}{#2}}%
    {\@fmsl@sh\textstyle{#1}{#2}}%
    {\@fmsl@sh\scriptstyle{#1}{#2}}%
    {\@fmsl@sh\scriptscriptstyle{#1}{#2}}}
\def\@fmsl@sh#1#2#3{\m@th\ooalign{$\hfil#1\mkern#2/\hfil$\crcr$#1#3$}}
\makeatother


\title{\large{\bf Quantum Mechanics on  Noncommutative Spacetime}}

\author{Xavier~Calmet\thanks{xcalmet@ulb.ac.be} and Michele~Selvaggi\thanks{selvaggi@clipper.ens.fr} \\
Service de Physique Th\'eorique, CP225 \\
Boulevard du Triomphe \\
B-1050 Brussels \\
Belgium 
}

\date{July, 2006}

\maketitle

\begin{abstract}
We  consider  electrodynamics  on a noncommutative spacetime using the enveloping algebra approach and perform a non-relativistic expansion of  the effective action. We obtain the Hamiltonian for quantum mechanics formulated on a canonical noncommutative spacetime.   An interesting new feature of quantum mechanics formulated on a noncommutative spacetime is an intrinsic electric dipole moment. We note however that noncommutative intrinsic dipole moments are not observable in present experiments searching for an EDM of leptons or nuclei such as the neutron since they  are spin independent. These experiments are sensitive to the energy difference between two states and the noncommutative effect thus cancels out.  Bounds on the noncommutative scale found in the literature relying on such intrinsic electric dipole moment are thus incorrect.
\end{abstract}


\newpage

\section{Introduction}

Gauge theories formulated on a canonical noncommutative spacetime have
recently received lots of attention when they were shown to appear in
a certain limit of string theory (see e.g. \cite{Seiberg:1999vs}).
This is however not the only motivation to consider a noncommutative
spacetime. A natural way to implement the notion of minimal length
\cite{minlength1} in gauge theories and gravitational theories is to
formulate these models on a noncommutative spacetime. The aim of this
paper is to reconsider the bounds on spacetime noncommutativity. We
shall be dealing with the simplest example one can think of: the
canonical noncommutative spacetime.

The idea of spacetime noncommutativity is not new and was first discussed by Snyder \cite{Snyder:1947qz}  in the early days of quantum field theory at a time where these theories were still plagued by infinities. The motivation to consider spacetime noncommutativity was that introducing a cutoff could help to deal with infinities. Nowadays we know that the quantum field theories relevant for the electroweak and strong interactions are renormalizable and thus cutoff independent, but modifying spacetime at short distance might be relevant for quantum gravity, whatever this theory might be. It is also well-known that non-commuting coordinates are relevant to nature, as soon as one restricts a system to the first Landau level. A textbook example is an electron in a strong magnetic field.

In this work we shall derive quantum mechanics  from an action for electrodynamics  formulated on a noncommutative spacetime taking the fields in the enveloping algebra. The case of Lie algebra valued fields has been treated in the literature \cite{Mezincescu:2000zq,Gamboa:2000yq,Chaichian:2000si,Nair:2000ii,Ho:2001aa,Chaichian:2002ew,Colatto:2005fy}. Our main motivation to study quantum mechanics using the enveloping algebra approach  \cite{Madore:2000en,Jurco:2000ja,Jurco:2001rq,Calmet:2001na}  is that this approach allows a formulation of the standard model on a noncommutative spacetime \cite{Calmet:2001na}. As we shall discuss in this work, spacetime noncommutativity introduces a new source of CP violation in noncommutative gauge theories. In gauge theories beyond the standard model, new sources of CP violation are typically giving rise to potentially large electric dipole moments. Electric dipole moments of the electron, muon, neutron and other nuclei are expected to be extremely tiny within the standard model. Even a tiny amount of CP violation coming from physics beyond the standard model could thus naively have a big impact on the electric dipole moments of these particles. Electric dipoles moment are low energy phenomena and the relevant experiments are performed at low energy. An non-relativistic limit of the noncommutative action for electrodynamics  is thus required to study these phenomena.

A canonical noncommutative
spacetime is defined by the noncommutative algebra
\begin{eqnarray} \label{NCA}
[ \hat x^\mu,\hat x^\nu]=i\theta^{\mu\nu}
\end{eqnarray}
where $\mu$ and $\nu$ run from 0 to 3 and where $\theta^{\mu\nu}$ is
constant and antisymmetric with mass dimension minus two. 
 We take $\theta^{\mu\nu}$ of the form:
 \begin{eqnarray}
(\theta^{\mu\nu})&=&
\left(
\begin{array}{cccc}
 0 & -C_1/c  & -C_2/c  & -C_3/c \\
 C_1/c & 0  & D_3  & -D_2 \\
 C_2/c & -D_3  & 0  & D_1\\
  C_3/c&  D_2 &  -D_1 & 0
\end{array}
\right)
\end{eqnarray}
where $c$ is the speed of light. Since we shall consider a non-relativistic limit, it is important to keep the factor $c$ explicitly. The vectors $\vec C$ and $\vec D$ are dimension full. In principle each of the components of these vectors could correspond to a different scale. We assume for the sake of simplicity that there is only one noncommutative scale which we denote by $\Lambda_{NC}$. We then have $|\vec C| \propto 1/\Lambda^2_{NC}$ and $|\vec D|\propto 1/\Lambda^2_{NC}$ . This definition is  analog to that of the field strength tensor of electromagnetism:
 \begin{eqnarray}
(F_{\mu\nu})&=&
\left(
\begin{array}{cccc}
 0 & -E_1/c  & -E_2/c  & -E_3/c \\
 E_1/c & 0  & B_3  & -B_2 \\
 E_2/c & -B_3  & 0  & B_1\\
  E_3/c&  B_2 &  -B_1 & 0
\end{array}
\right).
\end{eqnarray}

Formulating Yang-Mills theories relevant to particle physics on
such a spacetime requires to consider matter fields, gauge fields and
gauge transformations in the enveloping algebra otherwise SU(N) gauge
symmetries cannot be implemented.  The bounds on the noncommutative scale $\Lambda_{NC}$ relevant for the enveloping algebra approach \cite{Madore:2000en,Jurco:2000ja,Jurco:2001rq,Calmet:2001na}  are only fairly weak and of the order of a few TeVs \cite{Calmet:2004dn}. In this paper we shall show that the new  CP violation introduced through $\theta^{\mu\nu}$ in gauge theories formulated on noncommutative spaces does not lead to a tighter limit on the noncommutative scale despite a different  claim in the literature \cite{Chaichian:2002ew}. We will consider  low energy experiments such as  experiments searching for an intrinsic electric dipole moment  of leptons or nuclei such as the neutron. 

We first derive quantum mechanics on a noncommutative spacetime in the first section, then discuss the bounds on noncommutative intrinsic electric dipole moments from low energy experiments in section 2.  Finally, we conclude in section 3.

\section{Quantum Mechanics on Noncommutative Spacetime}

The enveloping algebra approach \cite{Madore:2000en,Jurco:2000ja,Jurco:2001rq,Calmet:2001na} allows to map a noncommutative action $\hat S$ on an effective action formulated on a regular commutative
spacetime. The dimension four operators are the usual ones and the
noncommutative nature of spacetime is encoded into  higher order
operators. In this section we will be considering electrodynamics on a noncommutative spacetime. The noncommutative action for a Dirac fermion
coupled to a U(1) gauge field is given by
\begin{eqnarray} \label{ncqed}
\int d^4x \bar{\hat \Psi}(\hat x) (i\hat{\fmslash D} -m c) \hat \Psi(\hat x) -\frac{1}{4} \frac{e^2}{c^2} \int d^4x \hat F_{\mu\nu} (\hat x) 
\hat F^{\mu\nu} (\hat x)
\end{eqnarray}
where the hat on the coordinate $x$ indicates that the functions belong
to the algebra of noncommutative functions and the hat over the
functions that they are to be considered in the enveloping algebra. Throughout this paper will shall keep the speed of light $c$ explicitly in our calculations, but we set $\hbar=1$.
The procedure \cite{Madore:2000en,Jurco:2000ja,Jurco:2001rq,Calmet:2001na}
to map actions such as (\ref{ncqed}) on an effective
action formulated on a commutative spacetime requires to first define
a vector space isomorphism that maps the algebra of noncommutative
functions on the algebra of commutative functions. The price to pay to
replace the noncommutative argument of the function by a commutative
one is the introduction of a star product: $f(\hat x) g(\hat x) = f(x)
\star g(x)$.  It turns out that this theory is renormalizable \cite{Calmet:2006zy} and invariant under noncommutative Lorentz transformations \cite{Calmet:2004ii}. One then expands the fields in the enveloping algebra
using the Seiberg-Witten maps \cite{Seiberg:1999vs} and obtains:
\begin{eqnarray}
\label{actionqed}
\int  \bar{\hat\Psi}(\hat x)  (i  \fmslash{\hat D}-m c) \hat \Psi (\hat x) d^4x 
 &=&
 \int  \bar{\psi} (i   \fmslash{D}- m c)\psi d^4x  \\ 
 \nonumber 
 && +\frac{1}{4} \frac{e}{ c} \int 
\theta^{\mu \nu}
 \bar{\psi} F_{\mu \nu} (i  \fmslash{D} -m c)\psi d^4x 
\\  \nonumber 
 && 
+\frac{1}{2}  \frac{e}{ c}  \int \theta^{\mu \nu}
 \bar{\psi} \gamma^\rho F_{\rho \mu} i  D_\nu \psi  d^4x 
\\  
-\frac{1}{4} \int \hat F_{\mu \nu } (\hat x)
 \hat F^{\mu \nu}(\hat x) d^4 x&=&-\frac{1}{4}  \ \int  F_{\mu \nu }
 F^{\mu \nu } d^4x
 \\ \nonumber &&
 -\frac{1}{8}   \frac{e}{ c}\
\int \theta^{\sigma \rho } F_{\sigma \rho }F_{\mu \nu } F^{\mu \nu } d^4x 
\\ \nonumber &&
+\frac{1}{2} \frac{e}{ c}\  \int  \theta^{\sigma \rho } F_{\mu \sigma}F_{\nu \rho} F^{\mu \nu} d^4x,
\end{eqnarray} 
to  first order in $\theta^{\mu\nu}$ and where as usual $F^{\mu\nu}=\partial^\mu A^\nu - \partial^\nu A^\mu$ and $A^\mu=(\Phi,A^i)$.

Using eq. (\ref{actionqed}), it is easy to derive the noncommutative Dirac equation, we find:
\begin{eqnarray}
(i\fmslash{D} - m c ) \psi +\frac{e}{4 c} \theta^{\mu\nu} F_{\mu\nu} (i\fmslash{D} - m c ) \psi  +\frac{1}{2}
\frac{e}{c} \theta^{\mu\nu} \gamma^\rho F_{\rho \mu} i D_\nu \psi=0,
\end{eqnarray} 
performing a non-relativistic expansion of this equation and splitting the bi-spinor $\psi$ into two components to prepare the non-relativistic expansion:
\begin{eqnarray}
\psi=\left(
\begin{array}{c}
  \tilde \phi   \\
  \tilde \chi   
\end{array}
\right)= e^{imc^2 t} 
\left(
\begin{array}{c}
  \phi   \\
  \chi   
\end{array} \right)
\end{eqnarray}
we find two coupled differential equations:
\begin{eqnarray}
i \left ( 1 + \frac{e}{4 c} \theta^{\mu\nu} F_{\mu\nu} \right) \partial_t 
\left (
\begin{array}{c}
  \phi   \\
  \chi   
\end{array} \right )
&=& \left ( 1 + \frac{e}{4 c} \theta^{\mu\nu} F_{\mu\nu} \right) 
\left (
c \vec \sigma \cdot \vec \pi \left (
\begin{array}{c}
  \chi   \\
  \phi   
\end{array} \right )
+ e \Phi \left (
\begin{array}{c}
  \phi   \\
  \chi   
\end{array} \right )
- 2 m c^2 \left (
\begin{array}{c}
  0   \\
  \chi   
\end{array} \right )
\right) \nonumber \\ &&
-\frac{e}{2} \theta^{\mu 0} F_{0\mu}\left (m c^2 + i \partial_t -\frac{e}{c}
 \Phi \right) 
\left (
\begin{array}{c}
  \phi   \\
 - \chi   
\end{array} \right )
 \nonumber \\ &&
 +\frac{e}{2} \theta^{\mu 0} F_{k \mu} \sigma^k ( m c^2 + i \partial_t -
\frac{e}{c} \Phi )  \left (
\begin{array}{c}
  \chi   \\
 \phi   
\end{array} \right )
 \nonumber \\ &&
 - \frac{e}{2} \theta^{\mu j}F_{0\mu} \pi_j   \left (
\begin{array}{c}
  \phi   \\
 -\chi   
\end{array} \right )
+\frac{e}{2} \theta^{\mu j}F_{k \mu} \sigma^k \pi_j  \left (
\begin{array}{c}
  \chi   \\
 \phi   
\end{array} \right ),
\end{eqnarray}
within our approximation, i.e. leading order expansion in $\theta$, first relativistic correction and weak field limit we recover the usual relation between the small and large components
\begin{eqnarray}
\chi=\frac{\vec \sigma \cdot \vec \pi}{2 m c } \phi.
\end{eqnarray}
We obtain
\begin{eqnarray}
i\partial_t \phi &=& \left ( \frac{\vec \sigma \cdot \vec \pi \vec \sigma \cdot \vec \pi 
}{ 2 m }  + e \Phi\right) \phi  \\ \nonumber &&
-\frac{e}{2} \theta^{i 0}F_{0i} m c^2 \phi + \frac{e}{4 c} \theta^{i 0} F_{ki} \sigma^k \vec \sigma \cdot \vec \pi \phi - \frac{e}{2} \theta^{i j} F_{0i} \pi_j \phi + \frac{e}{4mc} \theta^{\mu j} F_{k\mu} \sigma^k \pi_j \vec \sigma \cdot \vec \pi \phi 
\end{eqnarray}
and
\begin{eqnarray}
i\partial_t \phi &=& \left ( \frac{p^2}{2 m} - \frac{e}{2 m c} (\vec L + 2 \vec S) \cdot \vec B + e\Phi \right) \phi 
 \\ \nonumber &&
-\frac{e}{2} \theta^{i0}F_{0i} m c^2 \phi + 
\frac{e}{4 c} \theta^{i0} F_{ki} \pi^k \phi
+ i
\frac{e}{c} S_j  \epsilon^{jkn} \theta^{i0} F_{ki} \pi_n  \phi
 - \frac{e}{2} \theta^{\mu j} F_{0\mu} \pi_j \phi \\ \nonumber &&
  +  \frac{e}{4mc} \theta^{\mu j} F_{k\mu} \pi_j \pi^k \phi
  + i \frac{e}{2mc} \sigma_l \epsilon^{lmn} \theta^{\mu j}F_{m\mu} \pi_j \pi_n \phi,
    \end{eqnarray}
where $S^i=\sigma^i/2$ is the spin operator. Dropping terms which are not linear in the fields, we obtain the  low energy Hamiltonian for quantum mechanics on a noncommutative spacetime:
\begin{eqnarray}
{\cal H}&=& \left ( \frac{p^2}{2 m} - \frac{e}{2 m c} (\vec L + 2 \vec
 S) \cdot \vec B +e \Phi \right ) \\ \nonumber && -\frac{e m}{2} \vec
 C\cdot \vec E -\frac{e}{4 c^2} \vec C \cdot \vec B \times \vec p - i
 \frac{e}{c^2} (\vec C \cdot \vec p\ \vec S \cdot \vec B - \vec S
 \cdot \vec C \ \vec B \cdot \vec p) - \frac{e}{2c} \vec E \cdot (\vec
 D \times \vec p) \\ \nonumber && + \frac{e}{4mc^3} \vec C \cdot \vec
 p \ \vec E \cdot \vec p + \frac{e}{4mc} (\vec D \cdot \vec p \ \vec B
 \cdot \vec p- \vec D \cdot \vec B \ \vec p \cdot \vec p) \\ \nonumber
 && + i \frac{e}{mc^3} (\vec C \cdot \vec p)\ \vec S \cdot ( \vec E
 \times \vec p) + i \frac{e}{mc} \vec S \cdot (\vec D \times \vec p) \
 \vec B \cdot \vec p. \\ \nonumber &=& \left ( \frac{p^2}{2 m} -
 \frac{e}{2 m c} (\vec L + 2 \vec S) \cdot \vec B \right) -
 \frac{e}{2c} \vec E \cdot (\vec D \times \vec p) \\ \nonumber && +
 \frac{e}{4mc} (\vec D \cdot \vec p \ \vec B \cdot \vec p- \vec D
 \cdot \vec B \ \vec p \cdot \vec p) + i \frac{e}{mc} \vec S \cdot
 (\vec D \times \vec p) \ \vec B \cdot \vec p,
\end{eqnarray}
where we have dropped the constant term $-\frac{e m}{2} \vec C\cdot
\vec E$ and terms of higher order in the non-relativistic expansion in
$1/c$. We shall discuss the meaning of these new noncommutative
operators in the next section. We note that the one loop contribution
to the electric dipole moment \cite{Calmet:2006zy} is also of the form
$ \vec E \cdot (\vec D \times \vec p)$.

\section{Noncommutative Electric Dipole Moments}

Spacetime noncommutativity introduces a new source of CP violation in the standard model. Let us first discuss the transformation properties of $\theta^{\mu\nu}$ under CP transformations. We have CP$(\vec D)=-\vec D$ and CP$(\vec C)=\vec C$. Furthermore as usual CP$(\vec B)=\vec B$, CP$(\vec E)=-\vec E$, CP$(\vec p)=-\vec p$, CP$(\vec S)=\vec S$ and  CP$(\vec L)=\vec L$.  In a similar manner we have  T$(\vec D)=-\vec D$, T$(\vec C)=\vec C$,   T$(\vec B)=\vec B$, T$(\vec E)=-\vec E$, T$(\vec p)=\vec p$, T$(\vec S)=\vec S$ and  T$(\vec L)=\vec L$ under time reversal. The transformation under charge conjugate of $\theta^{\mu\nu}$ is chosen in such a way that it is CPT invariant, i.e. C$(\theta^{\mu\nu})=-\theta^{\mu\nu}$. We can then see easily that the operators
 \begin{eqnarray}
&&- \frac{e}{2c} \vec E \cdot (\vec D \times \vec p) \\ 
 &&+ \frac{e}{4mc} (\vec D \cdot \vec p \ \vec B \cdot \vec p - \vec D \cdot \vec B \ \vec p \cdot \vec p) 
 \nonumber
 \\
  &&+ i \frac{e}{mc} \vec S \cdot (\vec D \times \vec p) \ \vec B \cdot \vec p \nonumber
\end{eqnarray}
are CP violating. Note that CPT is conserved.
We will concentrate on one of these  operators
\begin{eqnarray}
 \vec E \cdot \vec d_{NC}=\frac{e}{2} \theta^{\mu j} F_{0\mu} p_j = \frac{e}{2c} \vec E \cdot (\vec D \times \vec p)
\end{eqnarray}
as it gives rise to a noncommutative instrinsic electric dipole moment (EDM) and could have observable effects.  This term is however not an usual EDM since it is spin independent i.e. not of the form $\vec E \cdot \vec S$.

This operator has been identified  in the Lie algebra approach as well \cite{Chaichian:2000si,Chaichian:2002ew}. Naively, one might think that the operator $\vec d_{NC}$ will give a contribution to the Lamb shift in the hydrogen atom \cite{Chaichian:2000si,Chaichian:2002ew}. However, a clean treatment of the hydrogen atom requires to solve the Schroedinger equation for two particles. This has been done in ref.  \cite{Ho:2001aa} where it is shown that the relative coordinate in the two body problem is actually commutative and since the potential is translation invariant one can eliminate completely the effects of spacetime noncommutativity at tree level from the hydrogen atom problem.

The operator $\vec d_{NC}$ could also naively give rise to an noncommutative intrinsic electric dipole moment for the leptons and for the quarks \cite{Chaichian:2002ew}. But,  although $\vec d_{NC}$ does violate CP, it does not involve the spin, as the usual intrinsic EDM does. Experiments which are searching for an EDM of the electron \cite{Regan:2002ta} or the nuclei such as the neutron \cite{Baker:2006ts} measure the energy difference between two spin states. Since our NCEDM is not sensitive to the spin, the noncommutative contribution cancels out. There is thus no bound on the noncommutative scale coming from EDM measurements despite the claim made in ref.  \cite{Chaichian:2002ew}.

\section{Conclusions}
We have derived quantum mechanics on a canonical noncommutative spacetime using a non-relativistic limit of the action obtained in the case of fields and gauge transformations valued in the enveloping algebra. We show that in this case, as in the Lie algebra case, there is a new source of CP violation in the effective action. It appears in the form of a noncommutative intrinsic electric dipole moment.

We then used these noncommutative intrinsic electric dipole moments to
study bounds on the noncommutative scale relevant for the enveloping
algebra approach. In this approach, noncommutative gauge theories have
minimal deviations with respect to regular gauge theories, and it thus
quite difficult to put a bound on the new scale involved in these
models.  We found that these electric dipole moments are not
measurable in experiments searching for an EDM of leptons or of the
neutron.  Therefore there is no obvious bound on spacetime
noncommutativity coming from low energy effects such as an EDM or the
Lamb shift. The bounds on spacetime noncommutativity remain weak as
argued in \cite{Calmet:2004dn}.  One could think that these electric
dipole moments could have a large contribution to the Z width, it is
however easy to show that the effect is largely suppressed by fermion
masses. The only valid bound which is not a test of Lorentz invariance
\cite{Carroll:2001ws} and thus a direct test of spacetime
noncommutativity is that coming from $Z\to \gamma \gamma$
\cite{Behr:2002wx}. The calculation performed in \cite{Behr:2002wx}
can be translated into a bound on the noncommutative scale of the
order of 1 TeV, it is however slightly model dependent since it
depends on the choice of the representation of the gauge fields in the
enveloping algebra.

\bigskip
\subsection*{Acknowledgments}
\noindent 
This work was supported in part by the IISN and the Belgian science
policy office (IAP V/27).

\bigskip


\end{document}